\documentclass[12pt,preprint]{aastex}
\usepackage{emulateapj5}
\usepackage{apjfonts}

\begin{document}

\submitted{submitted Aug 28, 2002}
\title{Measuring Cluster Peculiar Velocities
and Temperatures at cm and mm Wavelengths}

\author{Gilbert P. Holder}
\affil{School of Natural Sciences, Institute for Advanced
Study, Princeton NJ 08540}
\email{holder@ias.edu}

\begin{abstract}
We present a detailed investigation of issues related to the
measurement of peculiar velocities and temperatures using
Sunyaev-Zel'dovich (SZ) effects. 
We estimate the accuracy to which peculiar
velocities and gas temperatures of distant galaxy clusters 
could be measured.
With $\mu K$ sensitivity on arcminute scales at several frequencies
it will be possible to measure peculiar velocities to an accuracy
of $\sim$130 km\,s$^{-1}$ and gas temperatures to better than 1 keV. 
The limiting factor for the accuracy of $v_{pec}$ is the presence
of bulk motions within the galaxy cluster, even for apparently
relaxed clusters.  The accuracy of the temperature is mainly limited
by noise. These results are independent of redshift.
Such constraints can best be achieved with only three frequencies: one in
the Rayleigh-Jeans region ($\nu<40$ GHz), one near 150 GHz, and
the third at 300 GHz or higher. Measurements at the null of the thermal 
SZ effect are of marginal utility, other than as a 
foreground/background monitor. 

\end{abstract}

\keywords{cosmology: theory  
--- large-scale structure of the universe
--- cosmic microwave background 
--- galaxies: clusters: general
}

\section{Introduction}
\label{sec:intro}

Observations of the Sunyaev-Zel'dovich (SZ) effect 
\citep{sunyaev72} are currently at a mature stage. 
Highly significant detections are now routine and the next
generation of instruments is about to exploit the SZ effect to provide deep
surveys of galaxy clusters (for a recent review see 
Carlstrom {\em et~al.} 2002 \nocite{carlstrom02}). 
All current and near-future instruments 
are devoted to studies of the {\em thermal} SZ effect. There are a host
of yet more subtle distortions of the cosmic microwave background (CMB)
spectrum that contain a wealth of information related to the cluster's
peculiar velocity ($v_{pec}$) and the gas temperature 
($T$) of the intracluster medium.

In this paper we investigate the most significant of the subtler
distortions, namely those due to the line-of-sight motion (the kinetic
SZ effect) and the distortion of the spectrum due to relativistic
effects (the relativistic thermal SZ effect). These effects are
discussed in several reviews \citep{rephaeli95,birkinshaw99} and
are presented in great detail by several authors 
\citep{sunyaev80,rephaeli95,challinor98,sazonov98,itoh98,nozawa98b,molnar99,
dolgov01}. 

At sensitivities below $\sim$1 $\mu K$ some higher order 
effects could be important such as multiple scatterings and transverse
velocities. At such low levels, CMB anisotropies will
be also be a very difficult contaminant.
For simplicity, we restrict ourselves to the most significant effects.

The importance of using the SZ effect to its fullest potential comes
from its redshift independence. As a spectral distortion of the CMB, 
it redshifts along with the CMB and the amplitude of the distortion 
does not suffer from cosmological dimming. If there are clusters at
redshift $z \sim 2$ (as there should be in standard models of structure
formation) they will be very faint in X-rays, and perhaps undetectable
if the gas has been preheated by an early burst of star formation.
An independent probe of the gas temperature would be invaluable, and
preliminary steps along these lines are being made \citep{hansen02}.
Measurements of peculiar velocities would allow reconstruction
of the large scale density field on scales comparable to the horizon
\citep{dore02}.
A firm understanding of the large scale density inhomogeneities would
allow new tests of galaxy formation and provide a view of the 
evolution of structure.

In the next section we lay out the physical effects which allow such
powerful tests. In \S\ref{sec:freqs} we investigate the observing
frequencies which are best suited for such an investigation.
The details of the simulations and map-making methods are outlined
in \S\ref{sec:sims} and results are presented in \S\ref{sec:results}. We
close with a discussion of practical issues and a summary.

\section{SZ Effects}
\label{sec:sz}

CMB photons have roughly a 1\% probability of interacting with 
free electrons in the deep potential wells of galaxy clusters. Compton
scattering leads to an exchange of energy between the cool photons
and hot electrons, causing a distortion of the CMB spectrum. This
is known as the thermal SZ effect. Both the cross-section for scattering
and the electron energy distribution depend on the electron gas temperature,
providing a means of probing the gas temperature. The general form of
the thermal SZ effect is a temperature decrement at low frequencies and
an increment at high frequencies. The main effects of the gas temperature
are to slightly reduce the amplitude of the distortion (in a frequency
dependent way) and to shift the null of the distortion to a slightly
higher frequency.

The kinetic SZ effect arises from a bulk velocity of the galaxy cluster
along the line of sight. The scattered CMB photons essentially
pick up a slight redshift or blueshift from the peculiar velocity of
the cluster. The imprint on the CMB is a purely thermal distortion, i.e.,
the spectrum is exactly that of a blackbody at a slightly higher or lower
temperature.

All of these effects are very small. The typical energy exchange per
scattering for the thermal SZ effect is only $\sim 1\%$
(i.e., $kT/m_ec^2$) and only 1\% of
photons, on average, undergo scattering. 
Therefore, the cumulative effect relative
to the 2.73 K background is on the order of one part in $10^{4}$.
The kinetic effect is even smaller, since $v_{pec}/c$ is typically on the
order of $10^{-3}$, making the typical fractional energy exchange an order
of magnitude below that expected for the thermal effect. 
The distortions of the thermal SZ effect
due to the gas temperature are 
comparable in magnitude to the kinetic SZ effect. 
Importantly, all of these effects
have different spectral behavior, manifesting themselves differently
as a function of frequency. Figure \ref{fig:rel_shift} shows
some SZ effects as a function of frequency, both for temperature
and intensity. The dominant effect is clearly the non-relativistic
thermal SZ effect, showing the well-known decrement at low frequencies
and increment at higher frequencies. Note that the optical depth for the
lower temperature cluster has been scaled to provide the same net
Comptonization.  Both peculiar velocity  and the gas temperature shift
the spectrum, but the effect of the velocity on the temperature
is constant with frequency, while the gas temperature provides a
non-trivial spectral signature.
This could allow a well-designed experiment
with several observing frequencies to separate these effects and
measure the gas temperature and peculiar velocity.


\centerline{{\vbox{\epsfxsize=8cm\epsfbox{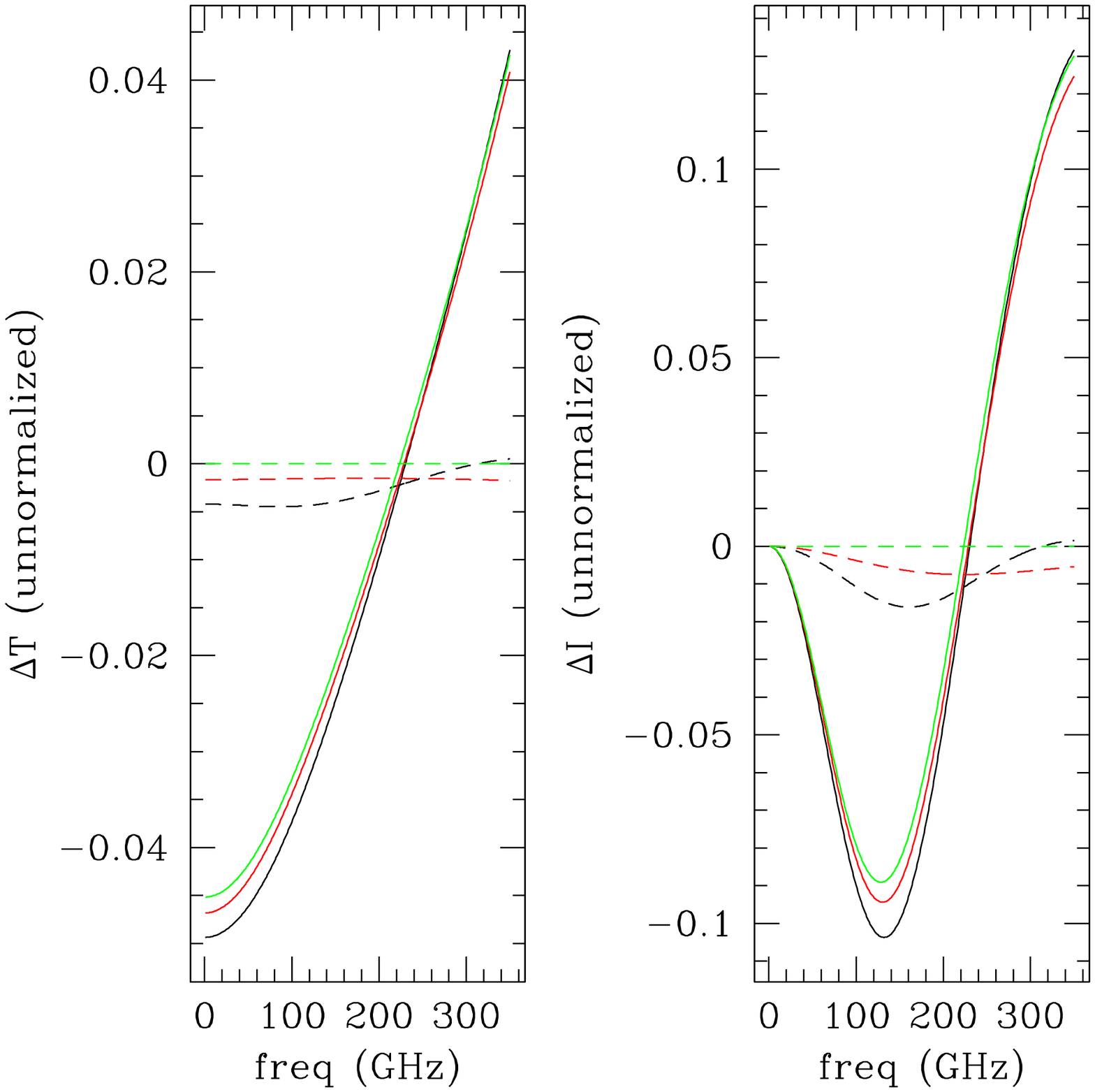}}}}
\figcaption{%
Effects of relativistic corrections and peculiar velocities
to SZ spectrum. Black curves show 
($T_e/keV,v/(km/s),\tau$)=(6,-500,0.02), red curves show
(12,-500,0.01), green shows (12,0,0.01). Dashed curves show all curves
relative to (12,0,0.01).
\label{fig:rel_shift}
 }%
\vspace{0.5cm}

\section{Observing Frequencies}
\label{sec:freqs}

Separation of the various SZ components requires at least three
observing frequencies (there are three unknowns: $\tau,T,v$).
To address the best positioning of these frequencies, a small
numerical experiment was performed. 

We take the range of observable frequencies to be 10-350 GHz, 
with lower frequencies likely contaminated by radio point sources
and higher frequencies possibly contaminated by galactic dust and
extragalactic dusty galaxies. In steps of 10 GHz, we assumed the
first frequency was at a frequency in the range 10-300 GHz. For
each of these values, we then stepped $\nu_2$ in 10 GHz intervals over all
frequencies from $\nu_1+10$ to 330 GHz, and then allowed the third 
frequency $\nu_3$ to be in the range $\nu_2+10$ to 350 GHz. 

At each position we use the fitting functions of \citet{nozawa98b}
to calculate the expected SZ temperature decrement or increment.
We use the Fisher matrix formalism to estimate errors in the three
parameters for each $(\nu_1,\nu_2,\nu_3)$ point. This method 
essentially approximates the likelihood function $\mathcal{L}$ as a Gaussian 
function of parameters $\vec{p}$ near 
its peak and assumes that the curvature (second derivative) at the 
peak gives a good estimate of uncertainties.

We define the Fisher matrix as
\begin{equation}
F_{ij} = < {d^2 \ln{\mathcal{L}} \over d p_i d p_j } > \quad ,
\end{equation}
where the angled brackets indicate an ensemble average over all realizations
of data. If the data products are taken to be $\Delta T_{\nu\alpha}$, the
temperature decrements or increments at frequencies $\nu=\nu_\alpha$, then the
likelihood of the parameters for a given set of data is simply 
\begin{equation}
-2\ln{\mathcal{L}} = \chi^2 = \sum_{\alpha=1}^3 
{[\Delta T_{\nu\alpha} - \Delta \bar{T}(\nu_\alpha,\vec{p})]^2 
\over \sigma_{\alpha}^2} \quad ,
\end{equation}
where $\Delta \bar{T}$ is simply the model prediction as a function of frequency
and fiducial model.

In this case, 
the Fisher matrix reduces to (after ensemble averaging)
\begin{equation}
F_{ij} = \sum_{\alpha=1}^3 {1 \over \sigma_{\alpha}^2 }
{ d \Delta \bar{T} \over d p_i}(\nu_\alpha,\vec{p})
{ d \Delta \bar{T} \over d p_j}(\nu_\alpha,\vec{p}) \quad ,
\end{equation}
where all derivatives are evaluated at the point where the $\vec{p}$ take
their fiducial values.


\centerline{{\vbox{\epsfxsize=8cm\epsfbox{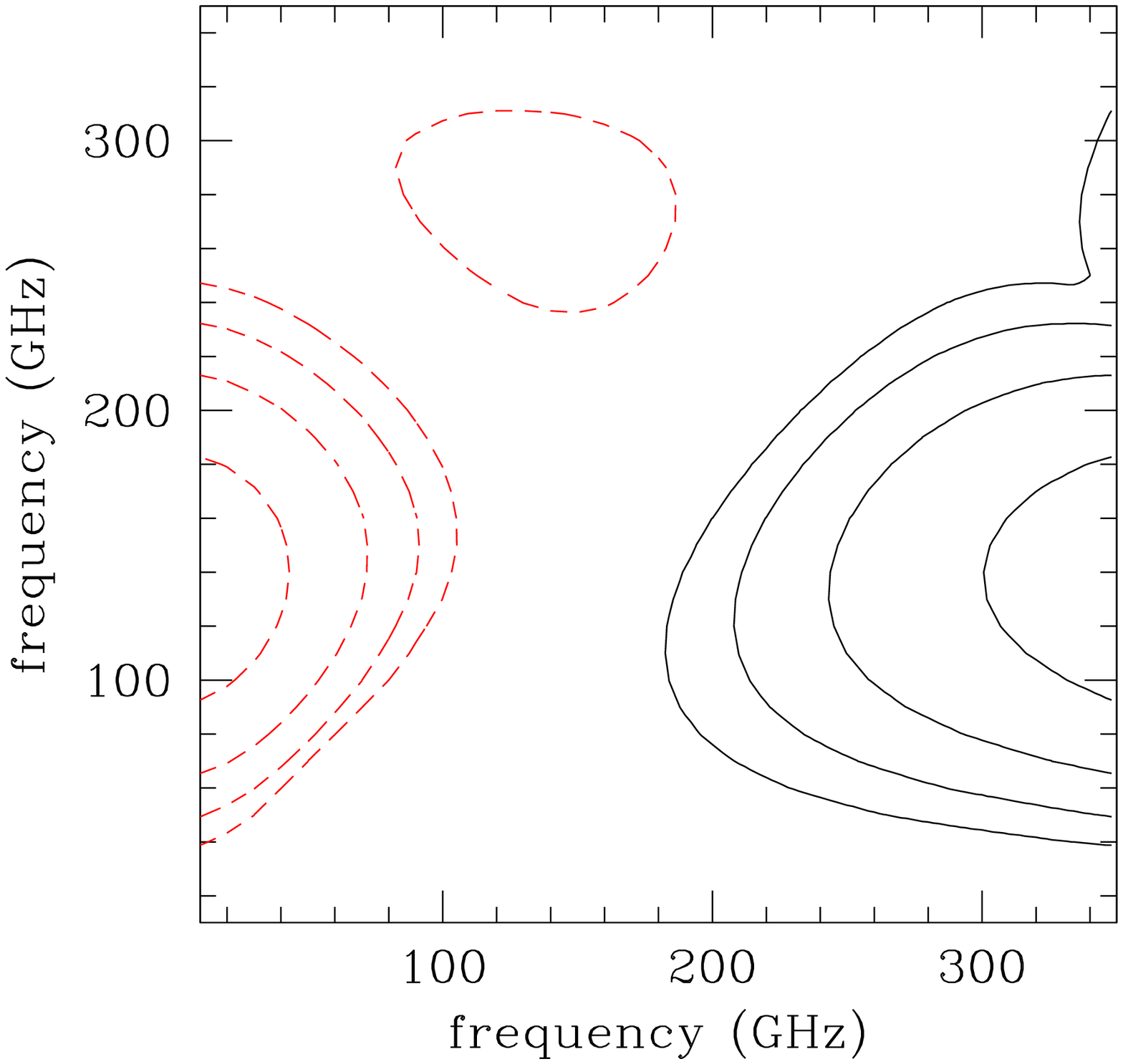}}}}
\figcaption{%
Uncertainty volumes as a function of observing frequencies. 
Red dashed contours show size of uncertainty volume for $\nu_1$ on
x-axis and $\nu_2$ on y-axis, while black contours instead are for
$\nu_3$ on the x-axis. Contours are in steps of 0.2 dex relative to
smallest volume. Fiducial model in this case has $T_e=12$ keV,
$v=$200 km/s, $\tau=0.01$.
\label{fig:dets}
 }%
\vspace{0.5cm}

Derivatives are calculated numerically by stepping from the fiducial
model by $\pm$0.1 keV in temperature, $\pm$10 km/s in velocity and
$\pm$5\% of the fiducial $\tau$.
The one sigma single parameter uncertainties, marginalized over other
parameters are then simply $(F^{-1})_{ii}$. Note that a Gaussian prior
in a parameter is easily applied by simply adding $1/\sigma_i^2$ to the
diagonal element $F_{ii}$, where $\sigma_i^2$ is the variance of the
prior on parameter $i$.

In figure \ref{fig:dets} we show the square root of the 
determinant of the inverse Fisher matrix as a function of frequencies,
assuming equal variances at each point of $\sigma_{\alpha}=1\mu K$. 
The square root of the determinant of the inverse Fisher matrix is a measure of
the volume of the ellipsoid in parameter space that  would give the 
single-parameter 68\% confidence regions. For uncorrelated parameters, the
determinant would simply be $\prod_{i=1}^3 \sigma_i^2$, where 
$\sigma_i^2$ is the variance in parameter $i$.

The best results are obtained 
if the first frequency is as low as possible and the third frequency is 
as high as possible. For practical purposes, all frequencies below about 
90 GHz work well, as do all frequencies above 300 GHz. What is interesting 
is that the second frequency is best placed at 150 GHz, not at the null 
of the thermal effect.  
Frequencies near the null of the thermal effect
($\sim$ 220 GHz) are not particularly good places to try to get 
constraints on parameters.  Furthermore, higher order terms to the
observed spectrum can become significant near the null, complicating
the analysis.

\centerline{{\vbox{\epsfxsize=9cm\epsfbox{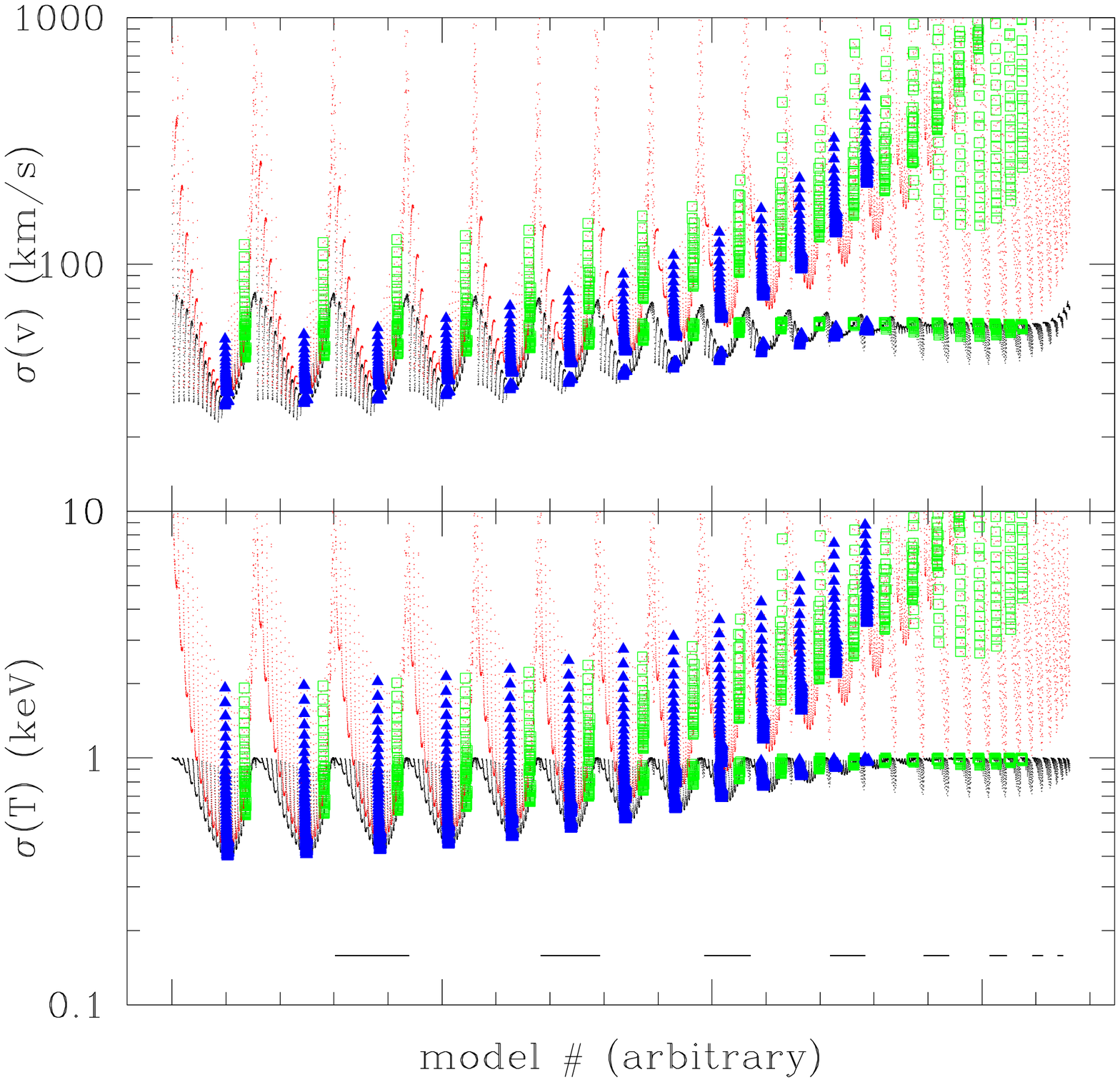}}}}
\figcaption{%
Possible measurement uncertainties for fiducial 1 $\mu K$
sensitivity experiment for different frequency coverages and
an assumed fiducial model of $T_e$=6 keV, $v=200$ km/s, $\tau$=0.01. 
Each model on the x-axis is a triplet of frequencies ($\nu_1,\nu_2,\nu_3$),
such that $\nu_1$ increases from 10 GHz at the left to 270 GHz at
the right (non-linearly). At each $\nu_1$ the other possible frequencies are 
similarly systematically checked. Red points are from SZ data alone
while black points show effect of a $\pm$ 1 keV independent measure
of the gas temperature. Horizontal lines in bottom panel indicate models
with $\nu_1$=30-240 GHz in steps of 30 GHz, i.e., every third
step in $\nu_1$. Open squares (green) indicate models with 
$\nu_2$=220 GHz while solid triangles (blue) show models with 
$\nu_2$=150 GHz.
\label{fig:sigs}
 }%
\vspace{0.5cm}

As a concrete example of why the thermal null is not necessarily a 
good place, take frequencies of 150, 220, 300 GHz. In this case, the
main effect of the temperature corrections is to reduce the thermal SZ
spectrum at the two end frequencies, and to shift the thermal SZ null to
a slightly higher frequency. Without an external temperature measurement,
this can be degenerate with a slightly lower Comptonization 
($\tau[kT/m_ec^2]$), which reduces the overall amplitude of the curve,
and a peculiar velocity away from the observer, which slightly shifts
the curve toward lower temperature decrements/increments. Whether or not
this degeneracy can be broken will depend somewhat on the fiducial model.

Note that we have assumed uniform errors at each frequency 
in {\em temperature} difference measurements (not intensity).  This is highly
idealized, in that foregrounds and noise are highly frequency-dependent.
As motivation for this choice, we assume that this sort of experiment
would be a natural outgrowth of a small-angle CMB experiment, where it
might be expected that the experimental goal would be something close
to uniform temperature sensitivity as a function of frequency. We
have also ignored finite bandwidth issues, which could be especially
problematic at the null of the thermal effect but should not be a large
problem at the frequencies which turn out to be optimal.

If one has gas temperatures from X-ray spectroscopy, the constraints on
parameters might be expected to improve significantly. We added a term
to the Fisher matrix corresponding to an independent gas temperature 
measurement of $\pm$1 keV. As shown in figure \ref{fig:sigs}
this does not significantly change the preferred frequency placement,
as can be seen by the lowest variances coinciding either with or without
external gas temperature information. Furthermore, at a sensitivity of
1 $\mu K$ the gas temperature measurement from the SZ data alone will be
sufficiently good that a $\pm$1 keV measurement is not of much
additional use at frequencies which are nearly optimal. For non-optimal
frequency coverage, the additional information on the gas temperature provides
tremendous leverage for velocity information.
From hydrodynamical simulations, it would not be expected
that the X-ray emission weighted temperature will agree with the
mean electron temperature (which is relevant here) to better than
1 keV \citep{mathiesen01}.

From the Fisher matrix estimates, single parameter uncertainties on gas
temperature, peculiar velocity, and optical depth, assuming a noise level
of $1 \mu K$ per frequency, are $\delta T \sim 0.5$\,keV, 
$\delta v \sim 30$km\,s$^{-1}$ and $\delta \tau \sim 6\times 10^{-4}$ for
a fiducial model of $T$=6 keV, $v_{pec}$=200 km\,s$^{-1}$, $\tau=0.01$.
These results will scale inversely with the assumed value of $\tau$
and will scale linearly with the assumed noise level.

\section{Simulations}
\label{sec:sims}

For an isothermal galaxy cluster with a unique peculiar velocity, the
results of the previous section are sufficient to estimate uncertainties.
In reality, however, galaxy clusters are not isothermal and bulk flows
on the order of the sound speed can persist for up to 10\% of the Hubble
time. 
In order to investigate the effects of such phenomena we turn to
numerical simulations of galaxy clusters.  

The simulations that we use 
have been described elsewhere \citep{holder00, mohr99, mohr97}. 
The cosmological model was a flat low density universe with matter
density (in units of the critical density) of 0.3 and a Hubble constant
of 80 km\,s$^{-1}$Mpc$^{-1}$.
The simulation method is P$^3$M-SPH \citep{evrard88}, 
where the hydrodynamics are done using the SPH method and the gravity is 
done with a particle-particle method for nearby particles and long-range 
forces are calculated from a grid. A dark matter only simulation was
performed in a large volume to identify the positions of galaxy clusters,
and all particles that ended up within the cluster virial radius were 
traced back to the initial conditions and replaced with more particles,
some dark matter and some gas, each of smaller mass. Large particles
outside the virial radius acted to include the effects of large scale
tidal fields. The net momentum of the simulation volume for each 
galaxy cluster was zero.

Each gas particle in an SPH scheme has a smoothing radius that roughly
corresponds to the mean distance to its Nth nearest neighbor, where
N is about 24. We made projected
maps of the relevant properties of the gas distribution (optical depth,
average velocity, gas temperature, mean Comptonization, SZ effects) along 
the three principal axes of the simulation by spreading the electrons 
associated with each gas particle uniformly within a disk of radius equal 
to roughly one third the SPH smoothing length. 
While this had the desirable effect of making the
maps smoother (visually) by reducing shot noise, the quantitative effects
of the smoothing kernel were negligible.

Three clusters were selected from the $z=0.5$ outputs of the simulations,
as the three most massive clusters in the simulation set at that redshift.
For each particle in the simulation, the gas temperature was used to
estimate its SZ effect contribution to the projected map,
again using the fitting functions of \citet{nozawa98b}. By adding up
the contribution of each particle we made synthetic SZ maps of each 
galaxy cluster, including the effects of relativistic corrections
and peculiar velocities.  The resulting maps had a resolution of
25''. The results are not sensitive to the resolution of the maps in
that higher resolution did not have any discernible effects. This
angular resolution also approximately corresponded to the spatial
resolution of the simulation.

At each point in the maps we used the SZ maps and the Fisher matrix
methods of the previous section to calculate the expected measurement
accuracy that would be possible in a high resolution, sensitive
experiment.  An angular diameter distance of 1000 Mpc was assumed.

\vskip 0.1in
\centerline{{\vbox{\epsfxsize=8cm\epsfbox{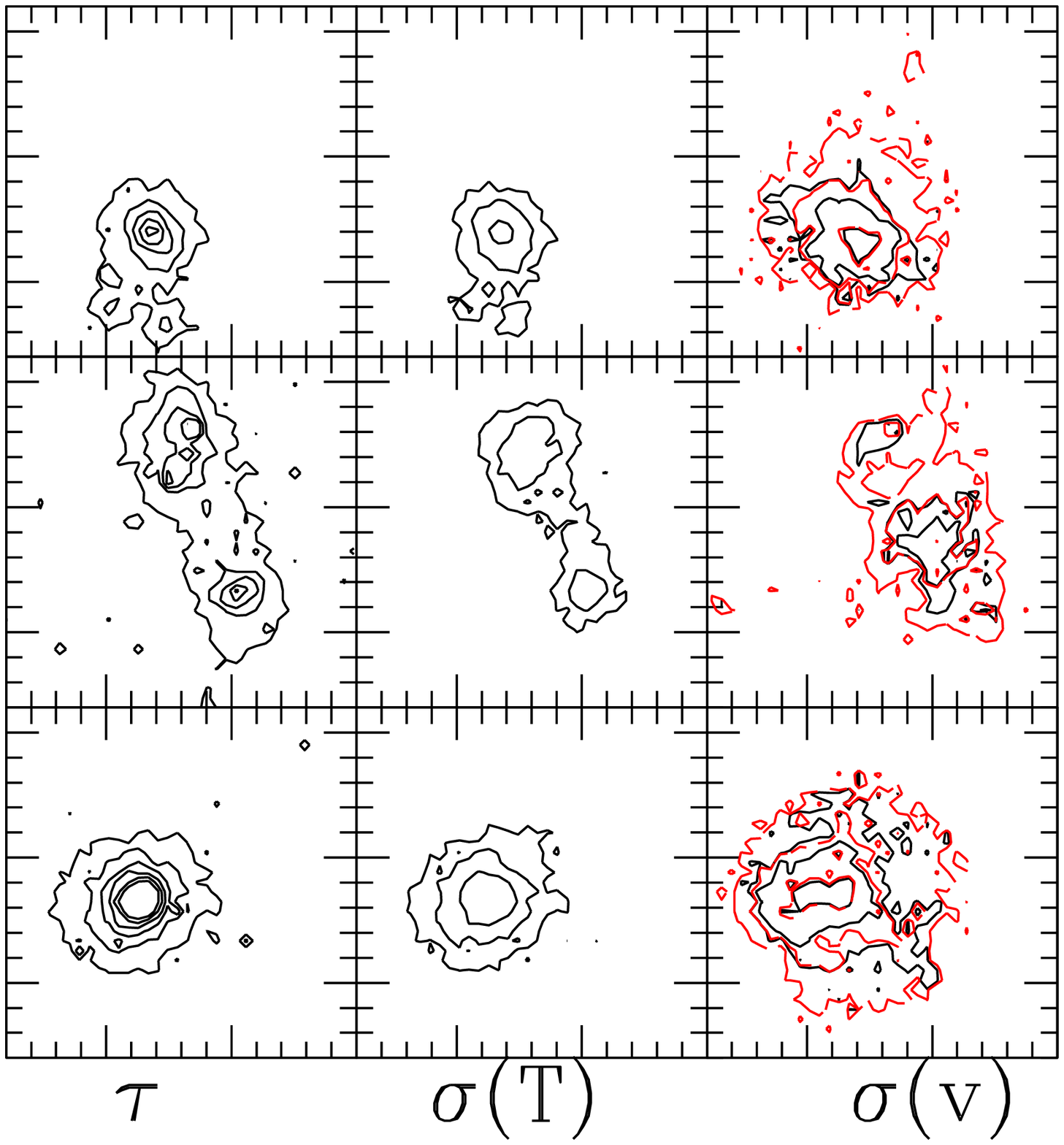}}}}
\figcaption{%
Surface maps of simulated galaxy clusters. From top to
bottom are three different simulated clusters.  From left to right
columns show maps of optical depth, possible measurement accuracies
on gas temperature and peculiar velocity. The maps are 14' on a side,
with each tickmark showing 1'. Contour levels in left panels
are at multiples of $2\times 10^{-3}$, in center panels correspond to
0.5, 1, 2 keV and in right panel show 20, 50 100 km/s.
Red dashed line in right panels correspond to measurement uncertainty 
in velocities assuming an independent measurement of the gas 
temperature of $\pm$1 keV.
\label{fig:maps}
 }%
\vspace{0.5cm}

\section{Results}
\label{sec:results}

As might be expected, the most important determinant of success in
parameter measurement is the optical depth, as shown in the simulation
maps in figure \ref{fig:maps}. All of the uncertainties
trace the optical depth map fairly well. In all clusters, within the
central few arcminutes the gas temperature can be measured to better
than 1 keV and the velocity to better than 100 km/s. Introducing
external gas temperature information improves the size of the 
region over which reliable velocity measurements are possible to roughly 7'.

The velocity uncertainty maps are much more ragged than the gas temperature
uncertainty maps. This is mainly due to ``lumpiness'' in the projected
velocity and gas temperature maps. In many ways, the velocity is the most
difficult quantity to measure, because it relies on a good measure of the
optical depth, which in turn rests on a reliable gas temperature measurement.
Being at the ``end of the line'' leads to increased sensitivity to 
small scale inhomogeneities.

Such accurate measures of the bulk velocity over scales of several 
arcminutes will in reality be virtually impossible. While the gas temperature
has a unique spectral signature, the effects of the bulk velocity have a
spectral signature that is identical to the intrinsic primary and
secondary anisotropies in the CMB. Most of the CMB anisotropy power is on much
larger angular scales, but it is instructive to estimate the region over which
one could best measure the bulk velocity.

\vskip 0.1in
\centerline{{\vbox{\epsfxsize=8cm\epsfbox{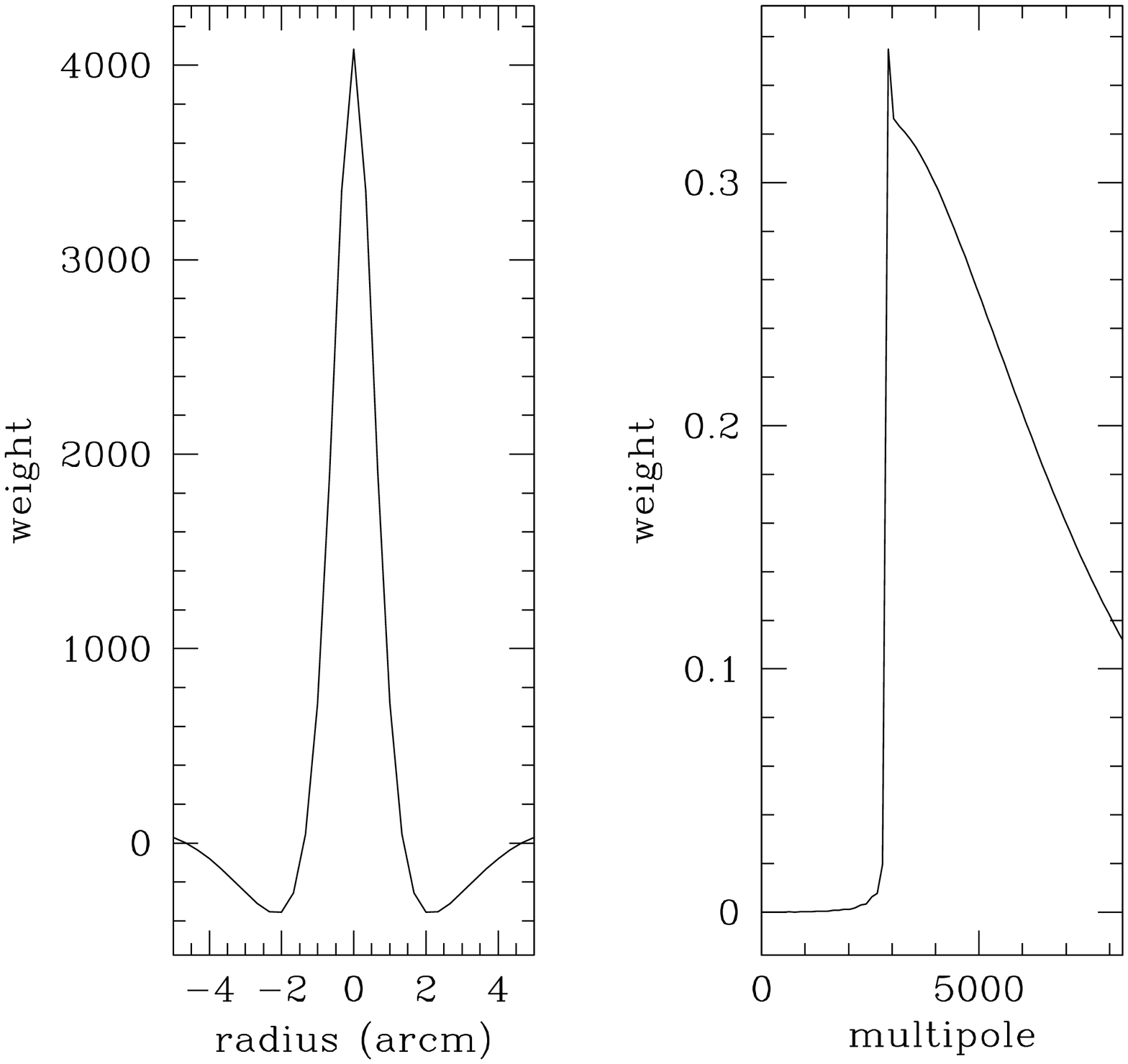}}}}
\figcaption{%
Filters necessary to remove noise from CMB primary and
secondary anisotropies in real space (left) and in Fourier space (right). 
The normalization is unimportant in this figure.
\label{fig:tegmark}
 }%
\vspace{0.5cm}

The tools for such an analysis are laid out in Haehnelt and
Tegmark (1996) \nocite{haehnelt96}. As an
update to their work, we investigate smaller scales and lower noise levels,
appropriate for the next generation of small-angle CMB experiments. We 
calculate the primary CMB anisotropies using the publicly available
CMBFast code \citep{seljak96}, and assume secondary anisotropies are
well described by flat band power and an rms temperature anisotropy 
of 1 $\mu K$,
and also assume thermal noise of 1 $\mu K$ in 1' pixels, although we ignore
beam effects. As a model for the optical depth, to be conservative we assume 
that the gas follows the 3D
density profile $\rho \propto [1+(r/r_c)^2]^{-1}$ with
$r_c=45''$ (expressed as an angle) and an isothermal gas. 
Clusters are almost certainly much more compact than this. The optimal
filter for such a profile is shown in figure \ref{fig:tegmark}. 
The Fourier space version is shown with spatial wavenumbers expressed
in units of multipoles, assuming a flat-sky approximation. To convert
to the usual $u-v$ plane of interferometry, $\ell= 2\pi R_{uv}$, where
$R_{uv}$ is the radius in the $u-v$ plane.  As can be
seen in the Fourier space profile, it is effectively a simple high-pass
filter that removes the large amount of power on large scales. From the
real space version, it can be seen that effectively all power beyond a
radius of 2' is filtered, with a FWHM of roughly 2'. The particulars
of the cluster model are essentially irrelevant, mainly affecting the
exact shape of the high $\ell$ taper and only modestly affecting the FWHM
of the filter. An excellent discussion of filtering issues is presented
in Haehnelt and Tegmark (1996).

What can be concluded from figure \ref{fig:tegmark} is that
peculiar velocity measurements of galaxy clusters can only
be performed in the inner two to four arcminutes because of severe 
contamination by CMB anisotropy. 
Because only a relatively small portion of the cluster is accessible
for observations, bulk velocities within the cluster, due to somewhat recent 
mergers can be a large source of confusion. Bulk flows moving at roughly the
sound speed (roughly 1000 km/s) containing ten percent of the mass
would be expected to mimic a signal of the entire cluster moving at
100 km/s. This is not a particularly unlikely occurrence, as a 10:1 mass 
ratio merger should happen quite regularly.  These bulk flows should
remain intact for a few crossing times. For a 1000 km/s flow crossing
1 Mpc, the crossing time is roughly 1 Gyr.

\centerline{{\vbox{\epsfxsize=8cm\epsfbox{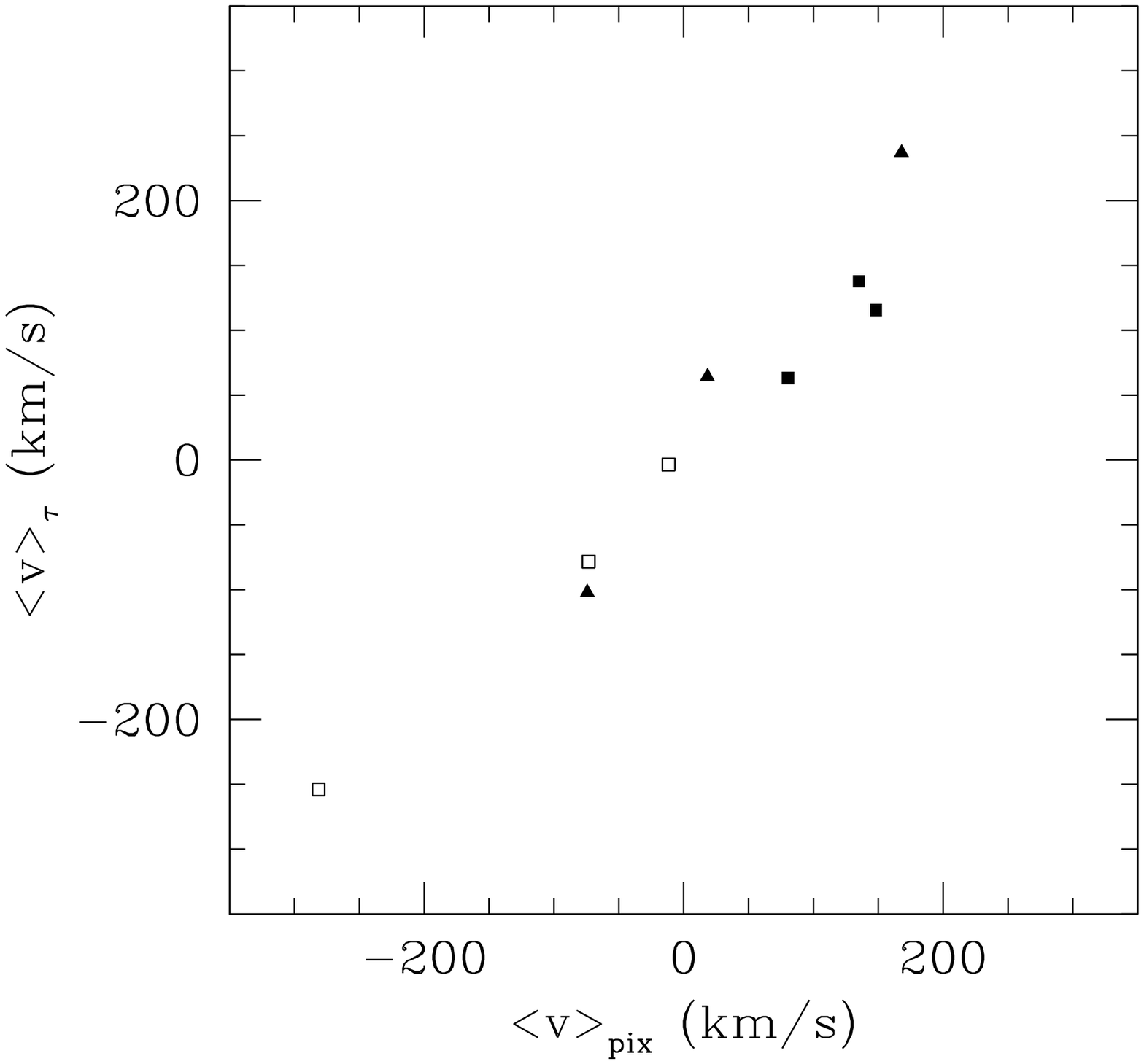}}}}
\figcaption{%
\label{fig:vels}
Comparison of average velocity in central 2' derived from
averaging the pixel values of the velocities vs averaging the pixels
weighted by the optical depth. Each simulated cluster has a different
symbol, and three projections of each cluster are shown (x,y,z). 
From top to bottom in figure \ref{fig:maps}: open squares, solid
triangles, solid squares.
 }%
\vspace{0.5cm}

In figure \ref{fig:vels} we show two measures of the effects of bulk flows on
peculiar velocity measurements using the simulated galaxy clusters. 
We calculated the average velocity within the
central two arcminutes of three projections of the three simulated clusters. 
This was done either
as an average of the individual pixels, or as an optical depth weighted
average of the pixels. A top-hat window was assumed, rather than the
optimal filter discussed above. For the simulation that is clearly a merging
system, the velocities were separately calculated for the central
regions of each clump and then averaged. The individual clumps for the 
merger event had individual velocities of $\sim$500 km/s in opposite 
directions. It would be readily apparent from the thermal SZ map that this
cluster is a double in the process of merging and it would be straightforward
to take this into account.
This should be a good estimate
of the typical errors that would be incurred due to bulk flows within the
intracluster medium. The dark matter in the simulation has no net momentum,
so a perfect reflection of the bulk velocity would be no net velocities.

From figure \ref{fig:vels}, it is apparent that there is significant
confusion coming from internal bulk flows. The {\em rms} velocities
from these three simulations are 135 km/s for the pixel-averaged velocity
and 141 km/s for the optical depth weighted average velocity. While it may
be possible to reduce this confusion somewhat, and more simulations are
required to quantify this more precisely, it is clear that measurements of
peculiar velocities to an accuracy better than 100 km/s will be very
difficult. This is in agreement with Haehnelt and Tegmark (1996), 
who found that bulk motions within a simulated galaxy cluster led to 
a $\sim$ 10\% mis-estimate of the input 1000 km/s peculiar velocity, 
suggesting that the bulk flows are contributing $\sim$100 km/s 
uncertainties. This has also been found independently by 
\citet{nagai02}. 

Because individual measurements will not be noise-limited, it may turn
out to be better to use the distribution of velocities rather
than velocities weighted by optical depth, in the same way optical studies use
galaxy velocity distributions to estimate the mean redshift. The coherence
length of the bulk flows are unfortunately not much smaller than a few
arcminutes, so it is not clear how much is gained.

\section{Discussion and Conclusions}
\label{sec:disc}

Measurements of peculiar velocities of galaxy clusters at microwave
frequencies will soon be possible. We have shown that multi-frequency,
sensitive observations could measure peculiar velocities
to an accuracy of roughly 100 km/s. 
Measurements of gas temperatures will be useful, with uncertainties
possibly smaller than could be achieved with X-ray spectroscopy. The redshift
independence of the SZ effect will make this an extremely powerful
tool for studies of distant clusters.
Exploring differences between X-ray emission weighted and SZ emission weighted
temperature maps will no doubt be interesting.

Foregrounds and backgrounds will be important barriers to such precise
studies of peculiar velocities. The primary anisotropies of the CMB
will become problematic on scales larger than a few arcminutes, and
point source removal  will be very difficult. It has not yet been demonstrated
that the atmosphere will not be a problem for ground-based experiments,
but there is no {\em a priori} evidence that there will be a problem.
Inter-frequency calibration to the requisite precision will be a
significant technical challenge.

Point sources come in (at least) two varieties. At frequencies below
$\sim$ 90 GHz radio point sources (primarily extragalactic AGN but also
star-forming galaxies) have
historically been a problem for SZ measurements and they are unlikely
to go away. The best solution seems to be simultaneous monitoring at
extremely high (a few arcseconds) resolution. At higher frequencies,
dusty star-forming galaxies are ubiquitous. If
no source subtraction is done, confusion could easily be at the level
of 10 $\mu K$, comparable to the kinetic SZ signal
\citep{blain98}. Currently very
little is known about these sources, making spectral subtraction (measuring
at a higher frequency where the SZ signal is negligible) difficult; 
the ultimate solution  may require something like ALMA
to remove the point sources at mm wavelengths. 
There is no evidence for variability in
these sources, so it will not be necessary to do the subtraction simultaneously,
as is required for the occasionally variable radio point sources.

The best frequencies for observation turn out to not include the
null of the thermal SZ effect. The best strategy is to have a Rayleigh-
Jeans band (below 90 GHz), a high frequency band (above 300 GHz) and
a band near 150 GHz.  Much has been made of the null of the thermal
SZ effect, and it will be important as a check for systematic errors,
but it is not a good frequency for cluster studies.

Bulk velocities within the cluster, combined with contamination from the
anisotropies of the CMB, lead to a limit of roughly 100 km/s on the possible
accuracy of kinetic SZ velocity measurements. This is much higher than
what would be expected from considerations of the background noise levels
and the 
few tens of km/s that arise from the difficulty in choosing the 
appropriate definition of velocity. More work on simulations could shed
significant light on optimal strategies for estimating the true peculiar
velocity as well as provide a much better estimate of the distribution
of errors that could be expected from an ensemble of galaxy cluster
peculiar velocity measurements.

Measurements at cm and mm wavelengths are opening a new window on cosmology.
It will soon be possible to measure gas temperatures and peculiar velocities
to good accuracy out to $z\sim 2$, 
allowing unprecedented tests of structure formation
as well as an excellent understanding of the topography of much of the
observable universe.

\acknowledgements{ I would like to thank Gus Evrard for generously
providing access to his simulations and for useful comments and 
clarifications.  This work has benefitted greatly from discussions with,
and encouragement from, Arthur Kosowsky, Erik Reese and Lloyd Knox.
GPH is supported by the W. M. Keck Foundation.}

{
}

\end{document}